\documentstyle[12pt,epsf]{article}
\begin{document}

\title{\vskip-2.5truecm{\hfill \baselineskip 14pt {{
\small  \\    
\hfill SISSA-55/97/EP \\
\hfill FTUV/97-19 \\
\hfill IFIC/97-19 \\ 
\hfill IC/97/38 \\
\hfill April 1997}}\vskip .1truecm}
 {\bf Finite temperature effects on }\\
{\bf CP violating asymmetries}}

\vspace{2cm}

\author{Laura Covi$^{1,2}$, Nuria Rius$^3$, Esteban
Roulet$^{3}$ and Francesco Vissani$^4$
 \\  \  \\
 $^{1}${\it International School for Advanced Studies, 
SISSA-ISAS},\\ {\it Via Beirut 2/4, I-34014, Trieste, Italy} \\ 
$^{2}${\it Istituto Nazionale di Fisica Nucleare, sez. di Trieste, Italy}\\ 
$^{3}${\it Depto. de F\'\i sica Te\'orica and IFIC, Centro Mixto }\\
{\it Universidad de Valencia-CSIC, Valencia, Spain}\\
$^4${\it International Centre for Theoretical Physics, ICTP,}\\ 
{\it Strada Costiera 11, Trieste, Italy}
}

\date{}
\maketitle
\vfill

\begin{abstract}
\baselineskip 20pt

We compute the $CP$ violating decay asymmetries relevant for
baryogenesis scenarios involving the out of equilibrium decays of
heavy particles, including the finite temperature effects arising from
the background of light thermal particles which are present during the
decay epoch. Thermal effects can modify the size of $CP$ violation by
a sizeable fraction in the decay of scalar particles, but we find
interesting cancellations in the thermal corrections affecting the
asymmetries in the decays of fermions, as well as in the decay of
scalars in supersymmetric theories. We also estimate the effects which
arise from the motion of the decaying particles with respect to
the background plasma.

\end{abstract}
\vfill
\thispagestyle{empty}

\newpage
\pagestyle{plain}
\setcounter{page}{1}
\baselineskip 24pt
\parskip 8pt

\section{Introduction}

A classic, and still very attractive, scenario for the generation of
the baryon asymmetry of the Universe, is based on the fact that very
massive  particles fall out of equilibrium as the temperature of the
Universe drops below their mass and their equilibrium density becomes
Boltzmann suppressed. If they decay at that epoch through baryon
($B$) violating channels which also violate $CP$, a net baryon
asymmetry will result. 

Usually, in these scenarios the $CP$ violation results from a
one--loop decay diagram, whose interference with the tree level
process allows the phases of the complex coupling constants to show up
in the decay asymmetries. 

In addition to the complex couplings, to have a non--zero partial
decay rate  asymmetry requires the loop integral to develop an
absorptive part. This actually happens whenever the loop diagram can
be cut in such a way that the particles in the cut lines can be
produced on--shell, and hence these particles need to be lighter than
the decaying one.
In the proposed baryogenesis scenarios of this kind, 
the light particles in the
loop are just standard quarks, leptons or Higgs bosons, so that the
appearance of a non--vanishing absorptive part is guaranteed. 
 
At the high temperatures at which the heavy particles decay, the light
standard particles are in equilibrium with the hot plasma present, and
hence a question arises on whether the existence of the background
particles has any effect in the evaluation of the $CP$ violating
asymmetries. Indeed, some time ago Takahashi \cite{ta84} showed that
the thermal effects could modify the predictions for baryogenesis in
$SU(5)$ models by up to $\sim 40\%$ with respect to the $T=0$ results,
and hence these effects may need to be taken into account in the
proper computation of the resulting baryon asymmetry in specific models.

The main effect of the background can be taken into account by employing
finite temperature propagators in the computation of the
loop. In this way it is possible to consider simultaneously both 
the `direct' propagation of a particle between two vertices in the loop
and the absorption by the medium of a particle from the first vertex
combined with the emission of another  one towards the second. These two
alternatives are actually indistinguishable in a thermal bath.

Another implication of the finite density background is the
modification of the final state phase space density distributions,
which take into account the stimulation of the decays into bosons and
the Pauli blocking of the decays into fermions, and this may
eventually also affect the rates.

\section{$SU(5)$ triplet decays}

In order to discuss these issues, let us start by reanalysing the
scalar decays into two fermions, which is relevant in the case of
heavy Higgs boson triplet decays in $SU(5)$. We will generalise the
computation in ref.~\cite{ta84} by including also the $CP$ violating
diagrams arising from mixing among different heavy states
\cite{ig79,bo91,li93,co96}. 
Notice that in order to have $CP$ violation at one loop in the $SU(5)$
GUT it is necessary to have more than one heavy state, and also the
existence of more than one heavy state is natural in other models,
such as the leptogenesis scenarios to be discussed later, and hence
the contribution from the mixing among the heavy states is generally
important.  
For simplicity we will consider the case in
which the masses of the heavy states are significantly splitted, since
the study of the $CP$ violation in the near degenerate situation
present additional complications \cite{li94,co97}. 

One generally expects that only the decay of the lightest of the heavy
states, $T_1$, will be the one leading to a net $B$, since any
asymmetry produced at earlier times through the decay of heavier
states would be erased by $B$ violating processes which could
still be in equilibrium\footnote{However, 
it is possible to imagine the situation in which 
a heavier triplet has the smallest decay rate, 
if the effect of the larger phase space 
is compensated by the smallness 
of the relevant couplings.}. 
Hence, one has to compute the asymmetry
resulting from the $T_1$ decay, and in the case we will consider in
which there is a hierarchy among the masses of the heavy states
($M_k\gg M_1$, with $k> 1$), it 
will be natural to assume that the heavier states $T_k$ have already
decayed, and hence have a negligible density, at temperatures $T\leq
M_1$ when the lighter
one is falling out of equilibrium.

Let us consider the $SU(5)$ lagrangian involving
several scalar five-plets $\Phi_i=(T_i, H_i),$ 
containing, together with Higgs doublets $H_i,$ the
heavy color triplets $T_i:$ 
\begin{equation}
{\cal L} =
f_i \Phi_{i\alpha}\  (\bar{\Psi}^{\alpha\beta}\ \chi_\beta)
+\frac{g_i}{8} \Phi _{i\alpha}\ 
(\epsilon^{\alpha\beta\gamma\delta\epsilon}\ 
\bar{\Psi}^c_{\beta\gamma}\ \Psi_{\delta\epsilon})
+h.c.,
\label{su5lag}
\end{equation}
where the gauge indices are denoted by greek letters.
The matter fields are in the decuplet and the fiveplet
representations as usual, $\Psi=(q,u^c,e^c)$ and
$\chi=(d,l^c).$ 
Since they are vectors in flavour space, the
 Yukawa couplings $f_i$ and $g_i$ should be thought as 
$3\times 3$ matrices, but for simplicity  
the flavour indices are not displayed.

The $CP$ violating $B$ asymmetry arising from the decay of a
$T_1$ and $\bar T_1$ pair is 

\begin{equation}
\epsilon={ \sum_f B_f
[\Gamma(T_1\to F_f)-\Gamma(\bar T_1\to \bar F_f)] \over
\sum_f [\Gamma(T_1\to F_f) + \Gamma(\bar T_1\to \bar F_f)]} ,
\label{eqeps}
\end{equation}
with $B_f$ the baryon number of the final states $F_f=\bar q\ell^c,\
\bar u e^c,\ \bar u^c d,\ \bar q^c q$.
{}From the interactions in
eq.~(\ref{su5lag}), at zero temperature we have
\begin{equation}
\epsilon={ 4 \sum_k\left[-{\rm Im}\!  \left\{{\rm Tr}\!  \left( g^\dagger_k 
g_1\, f_k f^\dagger_1  \right)\right\} 
{\rm Im}\!  \left\{  I_t(x_k)  \right\} +{1\over 1-x_k}
{\rm Im}\!  \left\{{\rm Tr}\!  \left( f_kf^\dagger_1 \right)
{\rm Tr}\!  \left( g^\dagger_k g_1   \right)  \right\} 
{\rm Im}\!  \left\{ I_s \right\} \right]\over 
 \left[ 
4 {\rm Tr}\!  \left(f_1 f^\dagger_1 \right)
+3 {\rm Tr}\!  \left( g^\dagger_1g_1\right) 	
       \right] },
\label{epsTriplets}
\end{equation}
where $x_k$ is the heavy-to-light ratio of squared masses:
\begin{equation} 
x_k\equiv \frac{M_k^2}{M_1^2} 
\label{xdef}
\end{equation}
Notice that the asymmetry is null if there is a single
triplet field. In this case, 
the result (\ref{epsTriplets}) depends on the 
hermitian matrices 
$g^\dagger_1 g_1$ and 
$f_1 f^\dagger_1;$ therefore 
their trace has no imaginary part,
and the trace of the product is equal 
to half the trace of the anticommutator, 
again an hermitian matrix. 
The terms in the denominator correspond to the 
tree level decay rate; those in the numerator to the interference 
between the tree level and the absorbptive part of the
loop amplitudes\footnote{The apparent difference with respect to refs. 
\cite{yi80,ta84} in the denominator is just due to a different normalization
of the field $\Psi$, and hence of the Yukawa couplings 
$f_i$. However, our result for the numerator differs from 
ref. \cite{bo91} in the sign of the first term (in agreement with 
\cite{na79,li93}) and a factor 2 in the second one.}. 
In particular, $I_{t}$ arises from the loop integral
in the `vertex' contribution, in which the $T_k$ is exchanged in the 
$t$--channel (see fig. 1a), and $I_{s}$ comes from the loop 
integral in the `wave' contribution, in which the $T_k$ is exchanged 
in the $s$--channel (fig.~1b), i.e.
\begin{equation} 
\bar u(p_1)v(p_2)I_t(x_k)=\bar u(p_1)\int{d^4q\over (2\pi)^4}
S(p-q,0)S(-q,0)D(q-p_2,M_k) v(p_2),
\end{equation}
\begin{equation}
M_1^2 I_s =-{i\over 2}\int {d^4q\over (2\pi)^4}
{\rm Tr}\!  \left(  S(-q,0)S(p-q,0)  \right),
\end{equation}
with $S(p,m)$ ($D(p,m)$) being the propagator of a fermion (scalar) 
with mass $m$ and momentum $p$.

{}From these expressions one gets the following absorptive parts: 
\begin{equation}
{\rm Im}\!  \left\{I_t(x ) \right\} ={1\over 16\pi}\left[1-x\ {\rm
ln}\left(1+{1\over x}\right)\right],
\end{equation}
 and
\begin{equation}
{\rm Im}\!  \left\{ I_s \right\} ={1\over 16\pi}.
\end{equation}

We consider now the finite temperature effects on the $CP$ violating
asymmetry $\epsilon$. The main effect comes from using instead
of the usual $T=0$ propagators, the finite $T$ ones.
We choose to work in the real time formalism (RTF) of thermal field 
theory \cite{rtf}. The Green's functions computed in this formalism 
are directly the time ordered ones, 
unlike in the imaginary time formalism where different analytical
extensions to real momenta lead to different Green's functions
(retarded, advanced, etc.) \cite{itf}. 
The RTF involves the introduction of a ghost field dual to each 
physical field, which leads to a doubling of the degrees of freedom. 
The thermal propagator has then a $2 \times 2$ matrix
structure: the (11) component refers to the physical field, 
the (22) component to the corresponding ghost field 
and the off-diagonal (12) and (21) components mixing them. 
However, since we are working only at one loop
and the external legs are physical, i.e. type--1 fields, 
we just need the (11) component of the propagators,  
which are, for fermions and bosons respectively,
\begin{equation}
S_{11}(p,m)=(\not\!p+m)\left[{i\over p^2-m^2+i0^+}-2\pi 
n_F(p\cdot u)\delta(p^2-m^2) \right],
\label{prof}
\end{equation}

\begin{equation}
D_{11}(p,m)=\left[{i\over p^2-m^2+i0^+}+2\pi 
n_B(p\cdot u)\delta(p^2-m^2) \right],
\label{prob}
\end{equation}
with $u$ the 4--velocity of the medium ($u=(1,0,0,0)$ in the medium rest
frame), and 
\begin{equation}
n_{F,B}(x)={1\over {\rm exp}\left(\, |x|\, /\, T \, \right)\pm 1}.
\end{equation}
We drop the (11) subindex in the propagators from now on.

As discussed previously, we neglect the background
density  of the heavy triplets
\footnote{Notice that the exchange of $T_1$ in the one--loop
diagrams does not contribute to $\epsilon$, so that only $k>1$ are
relevant.}  $T_k$ (assuming $M_k\gg M_1$), and as a first approximation
we assume that the decaying particle $T_1$ is at rest (particle motion
effects will be considered in section 4).
Using the well known property,
\begin{equation}
{1 \over x \pm i0^+} = {\cal P} \left( {1 \over x }\right) 
\mp i \pi \delta(x)
\end{equation}
we get for the absorptive part of the vertex loop integral
\footnote{We have checked that the same result can be obtained by 
applying the Cutkosky cutting rules at finite temperature in the real 
time formalism \cite{cutt}.}

\begin{equation}
{\rm Im}\!  \left\{ I_t^T(x_k) \right\} ={\rm Im}\!  \left\{
I_t(x_k)  \right\}  
\left[1-2 \bar n_F+2 \bar n_F^2 \right]-{x_k\over 8\pi}\int_{x_k}^\infty
{du\over u+1}n_F\left({M_1u\over 2}\right)  ,
\label{eqitt}
\end{equation}
where
\begin{equation}
\bar n_{F,B}\equiv n_{F,B} \left ({M_1 \over 2} \right).
\end{equation}
This is similar to the result in ref.~\cite{ta84}, except for the
relative sign between the $T=0$ part and the finite temperature
corrections, implying that the temperature effects tend to reduce
(instead of enhancing) the $CP$ violation. The integral term in the
r.h.s. of eq.~(\ref{eqitt}) arises from the absorption of particles
from the background which are energetic enough so as to put the
intermediate state $T_k$ on--shell and hence make it contribute to the
absorptive part. This term becomes then extremely small in the case in
which the heavy masses have a significant hierarchy, i.e. for $M_k\gg
M_1$, since the amount of background particles which are energetic
enough is Boltzmann suppressed.

For the finite temperature `wave' contribution we obtain 
\begin{equation}
{\rm Im}\!  \left\{ I_s^T \right\}={\rm Im}\!  \left\{ I_s \right\}
\left[1-2 \bar n_F+2 \bar n_F^2 \right],
\label{eqits}
\end{equation}
so that the temperature dependence is similar to the one in the
leading term of the vertex part. 
To give a quantitative idea of 
the effect, we notice that if the temperature 
is taken to be 1, 1/3 or 1/10 of the lightest triplet mass, $M_1$,
the overall factor in (\ref{eqits}) 
including the temperature
effects is 0.53, 0.70 or 0.99 respectively. 
The physical interpretation of this
effect is simple: due to the
thermal background, the two
light fermions exchanged in the loop are
subject to a Pauli blocking, 
and this leads to a reduction of the
amount of $CP$ violation.

In this scenario there is no effect on $\epsilon$ resulting from the
final state blocking, since this just leads to overall factors
$(1-\bar n_F)^2$ multiplying the rates, and hence these factors
cancel in the ratio in eq.~(\ref{eqeps}). Other thermal effects, such
as thermal masses or wave function renormalization, are higher order
in the coupling constants and hence we neglect them.

\section{Leptogenesis scenarios}

The $SU(5)$ model discussed in the previous section, in spite of being
the prototype for the `out of equilibrium decay' 
scenarios of baryogenesis, has
the drawback that it generates no net $B-L$ asymmetry (a
characteristic  of $SU(5)$), and hence the $B$ generation is
vulnerable to the anomalous $B$ violating processes of the Standard
Model \cite{ma83} (which only leave $B-L$ unaffected), with the
consequence that all asymmetries generated within this
model will be eventually erased.

A very interesting way out to this problem \cite{fu86,lgen} 
is based on the
generation of a lepton ($L$) asymmetry at early times, by the out of
equilibrium decay of heavy isosinglet neutrinos (the usual ones
appearing in see--saw models for neutrino masses 
and naturally present in GUT models such
as $SO(10)$).

In this section we discuss
temperature effects in this kind of models,
first under the minimal assumption that the standard model 
spectrum is enlarged to include 
heavy right-handed neutrinos, and then, considering the
supersymmetric extension of the model.

\subsection{Non-supersymmetric case} 
The interactions of the heavy neutrinos $N_i$, in the
 basis in which their
mass matrix is diagonal, are given by the following Lagrangian
\begin{equation}
{\cal L}=-\lambda_{ai}\epsilon_{\alpha\beta}\bar
N_i P_L\ell^\alpha_aH^\beta+h.c.
\end{equation}
where $\ell_a =(\nu_a,\ l^-_a)$ and $H =(H^+,\ H^0)$ are the lepton and
Higgs Standard Model doublets
($a=e,\mu,\tau$, $i=1,2,3$, and $\epsilon_{\alpha\beta}=
-\epsilon_{\beta\alpha}$, with
$\epsilon_{12}=+1$).

The complete $T=0$ $CP$ violating $L$ asymmetry was computed for this
model in ref.~\cite{co96}, resulting in
\begin{equation}
\epsilon
= 2 \sum_{k>1}{\cal I}_{k1}
\left[ 
{\rm Im}\!  \left\{  J_t(x_k)  \right\} + 
2 {\sqrt{x_k}\over x_k-1} {\rm Im}\!  \left\{ J_s \right\}
\right],
\label{eps-t0} 
\end{equation}
 where $x_k$ is defined analogously to eq. (\ref{xdef}) and 
\begin{equation}
{\cal I}_{k1}\equiv { {\rm Im}\!  \left\{\left(
\lambda^\dagger\lambda\right)^2_{k1}\right\}
\over(\lambda^\dagger\lambda)_{11}
}.
\end{equation}
Notice that, in full analogy with the $SU(5)$ case, a single 
right handed neutrino would be 
unable to generate any asymmetry.

The loop integrals are given in this case by
\begin{equation}
\bar u(p_1)P_Ru(p)J_t(x_k)=\bar u(p_1)P_R\int {d^4q\over
(2\pi)^4}S(q,0)S(q-p_2,M_k) D(p-q,0) P_Lu(p)
\end{equation}
and
\begin{equation}
\bar u(p_1)u(p)M_1J_s=-i\bar u(p_1)\int{d^4q\over 
(2\pi)^4}S(q,0)D(p-q,0)u(p),
\end{equation}
so that the absorptive parts result
\begin{equation}
{\rm Im}\!  \left\{  J_t(x)  \right\}=
{1\over 16 \pi}\sqrt{x}\left[1-(1+x){\rm
ln}\left(1+{1\over x}\right)\right] ,
\label{jt-func}
\end{equation}
and 
\begin{equation}
{\rm Im}\!  \left\{ J_s  \right\}=-{1\over 32\pi}.
\end{equation}

The computation of the finite temperature contribution to these
quantities is similar to the one in the previous section, and 
leads to
\begin{equation}
\begin{array}{l}
{\rm Im}\!  \left\{  J^T_t(x_k) \right\}=\begin{array}[t]{l} 
\displaystyle  {\rm Im}\!  \left\{  J_t(x_k) \right\}
\left[ 1-\bar n_F+\bar n_B-2 \bar n_F\bar n_B\right]\\[2ex]
+ \displaystyle  {\sqrt{x_k}\over 16\pi}
\int_{x_k}^\infty 
{du\over u+1}\left[n_F\left({M_1u\over 2}\right)(u-x_k)
+n_B\left({M_1u\over 2}\right)(x_k+1) \right],
\end{array} \\[8ex]
\displaystyle {\rm Im}\!  \left\{ J^T_s \right\}=
{\rm Im}\!  \left\{  J_s \right\}\left[
1-\bar n_F+\bar n_B-2 \bar 
n_F\bar n_B\right],
\end{array}
\end{equation}
where for instance in this last expression 
the different contributions in the r.h.s. clearly separate into
 the pieces coming from the $T=0$ propagators, the one from the
thermal correction to the fermion ($\ell$) propagator, the one from
the thermal piece of the
 boson ($H$) propagator and the product of these two corrections.
However, it is easy to check that the Bose and Fermi distributions
satisfy
\begin{equation}
n_B(E)-n_F(E)=2n_B(E)n_F(E),
\end{equation}
so that the main temperature correction cancels out (only the
small integral term in ${\rm Im}\!  \left\{J_t\right\}$ survives).
 This is due to the
opposite effects resulting from the Pauli blocking of the loop fermion
line and the stimulation of the bosonic loop line. On the other hand,
here again the final state statistical factors
$(1-\bar n_F)(1+\bar n_B)$ cancel in the expression for $\epsilon$,
and as a result no significant temperature dependent effect is found.

\subsection{Supersymmetric case}
This scenario has also received considerable attention within a
supersymmetric framework \cite{ca93,mu93,co96}, in 
particular because the scalar partner of
the heavy neutrino $\tilde N_1$ is a good candidate for being the
inflaton field, in which case the $L$ asymmetry could  be produced
during the process of reheating of the Universe as $\tilde N_1$ decays
\cite{mu93}. 
In this case, as we will now show, 
another interesting cancellation is found in the
asymmetry produced by $\tilde N_1$ decay.

The scalar neutrino can decay either into two scalars ($\tilde LH$)
or into two fermions ($\ell \tilde h$), and the contribution to $CP$
violation in one channel is obtained from the loop involving the
particles of the other channel \cite{co96}. 
The thermal effects will modify the
asymmetries corresponding to each channel, in a way similar as they
did in the case of $SU(5)$. For instance, in the $\tilde LH$ channel,
if we ignore the small integral piece coming from the vertex
(equivalent to the last term in the r.h.s. of eq.~(\ref{eqitt})), 
the asymmetry will be
\begin{equation}
\epsilon^T(\tilde N\to\tilde LH)\simeq
\epsilon^{T=0}(\tilde N\to\tilde LH)\left[1-2 \bar n_F(1-\bar
n_F)\right],
\end{equation}
and similarly
\begin{equation}
\epsilon^T(\tilde N\to\ell\tilde h)\simeq
\epsilon^{T=0}(\tilde N\to\ell\tilde h)\left[1+2 \bar n_B(1+\bar
n_B)\right].
\end{equation}

However, due to the effects of the final state phase space factors
$(1\mp \bar 
n_{F,B})$ entering into the partial decay rates, the branching ratios
of the two different channels will no longer be equal (as is the case
at $T=0$). One has instead that
\begin{equation}
BR(\tilde N\to\ell\tilde h)={(1-\bar n_F)^2\over (1-\bar
n_F)^2+(1+\bar n_B)^2 }=1-BR(\tilde N\to\tilde LH).
\end{equation}
The total asymmetry produced in the $\tilde N$ decay is
\begin{equation}
\epsilon^T_{\tilde N}=BR(\tilde N\to\ell\tilde h)\epsilon^T(\tilde
N\to\ell\tilde h) +BR(\tilde N\to\tilde LH)\epsilon^T(\tilde
N\to\tilde LH),
\end{equation}
with the surprising result that the main corrections arising from
thermal effects actually cancel out, leading to
\begin{equation} 
\epsilon^T_{\tilde N}\simeq\epsilon^{T=0}_{\tilde N}.
\end{equation}

Notice that there are also new supersymmetric diagrams contributing 
to the $CP$ violating asymmetry in the heavy neutrino decays.
However, since the particles in the loop as well as the external ones 
are always one fermion and one boson, both massless, the cancellation 
found in the previous subsection will also occur in the new
channels, and therefore there are no significant thermal 
corrections to the zero temperature result of ref. \cite{co96}. 

\section{Effects of particle motion}

In the discussion so far we have always considered the decay rate of a
particle at rest in the thermal bath. However, 
the decaying particle will in general be moving through the background
with non zero velocity $\vec{\beta}=\vec{v}/c$. 
Since now the Lorentz symmetry is
explicitly broken by the plasma, this motion 
can in principle affect the thermal
corrections to the decay asymmetry. When the leading
thermal corrections  cancel, as in the leptogenesis scenarios
discussed in Section~3, the effects of
the motion of the decaying particle will provide the
main thermal corrections, and hence it is worth to quantify them.
To estimate the size of these effects, we will consider here the case of 
the heavy neutrino decay in the non-supersymmetric model. 
The other cases can be analysed similarly.

It is convenient to compute the decay rate asymmetries in the
rest frame of the decaying particle, where the medium will be
characterised by a non trivial 4-velocity $ u = (\gamma,
 -\gamma\vec{ \beta})$, with $ \gamma = 1/\sqrt{1 - \beta^2} $ as
usual. 
In this system, the effect of the motion will reflect in a
modification of the equilibrium distributions appearing in the thermal
propagators, eqs.~(\ref{prof}--\ref{prob}). 
The decay rate will also depend on $\vec{\beta}$
through the final state phase space factors, which for the 
case of fermion decays is
\begin{equation}
\left[1-n_F(k_\ell\cdot u)\right]\left[1+n_B(k_H\cdot u)\right]=
{2\over 1-{\rm exp}(-M_1\gamma /T)}P_\beta(\cos\theta),
\end{equation}
where $\theta$ is the angle between ${\vec \beta}$ and 
the momentum of the final state
lepton,  and we have introduced
\begin{equation}
P_\beta(z)\equiv \frac{1}{2}
\left\{1-\sinh\left(\frac{z\beta \gamma M_1}{2T}\right)
\Big/\sinh\left(\frac{\gamma M_1}{2T}\right)\right\}^{-1}.
\label{denom}
\end{equation}
We see that now  there is a privileged direction selected by the plasma 
spatial velocity and therefore the decay process is no more isotropic:
the rate depends on the angle of the decay product trajectories with
respect to this direction. Due to statistics the fermions (bosons) are 
preferentially emitted in the direction parallel (antiparallel) to the 
plasma velocity, which corresponds to the less (more) occupied region 
of the thermal distribution. 
Anyway, since we are interested in the total decay rate, the angular 
dependence is integrated out and we are left only with a  $ \beta^2 $ 
dependence\footnote{The integrated decay
rate can depend only on the Lorentz invariants $ p^2 = M_1^2 $ 
and $ p \cdot u = M_1 \gamma $ \cite{we82}.}.

Let us define $\epsilon_\beta$ as the integrated 
asymmetry generated in the decay of
a heavy neutrino moving with
velocity $ {\bf \beta} $. 
The decaying particles will actually
have a distribution of velocities with occupation numbers $n(E)$,
where  $E=M_1\gamma$. To
estimate the overall effect of the particle motion we may just 
approximate this distribution 
with the Fermi Dirac distribution, 
$n(E)=n_F(E)$, and compute the average asymmetry 
\begin{equation}
\langle\epsilon\rangle={1\over N_1}\int {d^3p\over (2\pi)^3}
 \ n(\gamma M_1)\epsilon_\beta =\frac{1}{N_1}\int_0^1d\beta
\frac{dN_1}{d\beta}\epsilon_\beta, 
\label{mean-eps}
\end{equation}
where $N_1$ is the particle's volume density and 
\begin{equation}
\frac{dN_1}{d\beta}={M_1^3\over 2\pi^2} \beta^2 \gamma^5\ n(\gamma M_1).
\end{equation}
It is clear that $\langle\epsilon\rangle$, computed for a given
temperature $T$,  is just the asymmetry
that would result if the initial thermal population of heavy neutrinos
went out of equilibrium and decayed all simultaneously at temperature
$T$. We will use the asymmetry $\langle\epsilon\rangle$ as an
indicator of the possible effects of the particle motion, although
clearly to obtain the exact impact of this into the final lepton
asymmetry would require to integrate the whole Boltzmann equations, a
task beyond our scopes.

To obtain the decay asymmetry at a fixed velocity
$\epsilon_\beta$, it is convenient to separate the angular dependences
arising from the final state phase space factor $P_\beta(\cos\theta)$ and
the one arising from the one loop integrals (via the anisotropic
background density in the particle rest frame). This last will be
included in the factor $L_\beta(\cos\theta,x_k)$ defined through
\begin{equation}
\epsilon^c_\beta={2\over \int_{-1}^1dzP_\beta(z)}\sum_k{\cal I}_{k1}
\int_{-1}^1 dzL_\beta^c(z,x_k)P_\beta(z).
\label{vel-eps} 
\end{equation}
Here the supraindex $c$ labels the two different contributions to the CP
asymmetry in the decay, i.e. the ``wave" ($\epsilon^s$) and the ``vertex"
($\epsilon^t$)  pieces. 
Clearly at zero velocity we have, according to eq.~(\ref{eps-t0}), 
\begin{eqnarray}
L^s_0(z,x_k)&=&{2\sqrt{x_k}\over x_k-1}{\rm Im}\!  \left\{J_s\right\},\\
L^t_0(z,x_k)&=& {\rm Im}\!  \left\{J_t(x_k)\right\},
\label{jt}
\end{eqnarray}
which are actually independent of $z$ as expected.

 At $\beta\neq 0$ we find after direct computation of 
the interference terms 
\begin{equation}
L^s_\beta(z,x_k)=L^s_0(z,x_k)[f_\beta^{(0)}-zf_\beta^{(1)}],
\label{gamma-wave}
\end{equation}
where the loop functions $f_\beta^{(n)}$ 
\begin{equation}
f_\beta^{(n)}=\int_{-1}^1dy\ y^{n} \ P_\beta(y)
\label{loop-func}
\end{equation}
arise after integrating over the angle of the momenta in the loop.
Notice that the phase space factor $P_\beta$ appears also in these
integrals, arising from the statistical factors in the thermal
loops. 
In the limit of zero
velocity, we have $f^{(0)}_0=1$ and $f^{(1)}_0=0$.

Now we can write in a simple form the final expression for
$\epsilon_\beta^s;$
namely, putting together eqs.~(\ref{mean-eps}),
(\ref{vel-eps}) and (\ref{gamma-wave}) we get:
\begin{equation}
\langle \epsilon^s \rangle = \epsilon_0^s \frac{1}{N_1}
  \int_0^1  d\beta \  \frac{dN_1}{d\beta} \ 
f_\beta^{(0)}\left[ 1 -
\left(\frac{f_\beta^{(1)}}{f_\beta^{(0)}}\right)^2  \right].
\label{eps-wave}
\end{equation}

It is not possible to integrate analytically this expression,
but expanding the loop functions
to first order in  $ \beta^2 $  we obtain
\begin{eqnarray}
f_\beta^{(0)}&\simeq& 1 + \frac{1}{3}\  \beta^2 \
\left(\frac{M_1/2T}{\sinh{(M_1/2T)}}\right)^2,\\
f_\beta^{(1)}&\simeq& \frac{1}{3}\  \beta \ \frac{M_1/2T}{\sinh{(M_1/2T)}}. 
\end{eqnarray}
Substituting in eq.~(\ref{eps-wave}), we get
\begin{equation}
\langle\epsilon^s \rangle \simeq \epsilon_0^s \left[ 1 + \frac{2}{9}
\langle\beta^2  \rangle
\left(\frac{M_1/2T}{\sinh{(M_1/2T)}}\right)^2 \right].
\label{mean-eps-w}
\end{equation}

For $ M_1/T = 1,3, 10$ we have $ \langle\epsilon^s
\rangle/\epsilon_0^s  \simeq 1.168,
1.057, 1.000$; we see therefore that the effect is small compared to 
the $ \beta = 0 $ piece. In the previous evaluation we have used for
the average velocity just a simple approximation, which  is accurate
to the 10\%
level, writing $\langle\beta^2\rangle\simeq\langle\beta\rangle_{Boltz}^2$,
with the average velocity for a Boltzmann distribution of massive
relativistic particles being (with $x\equiv M/T$)
\begin{equation}
\langle\beta\rangle_{Boltz}={2(1+x){\rm exp}(-x)\over x^2K_2(x)}.
\end{equation}
Computing numerically eq.~(\ref{eps-wave}) we find, for the same reference
values of $M_1/T$ as before, that $ \langle\epsilon^s
\rangle/\epsilon_0^s = 1.231, 1.054$ and 1.000, so that the approximate 
result is acceptable.

Let us now consider the ``vertex" contribution. In this case the loop
integration is more involved and we get the following expression
(neglecting the small integral piece which is Boltzmann suppressed):
\begin{equation}
L^t_\beta(z,x_k)={\sqrt{x_k}\over 16\pi}
\left[ f_\beta^{(0)} - g_\beta (x_k,z) \right] ,
\label{l-vertex}
\end{equation}
where 
\begin{equation}
g_\beta (x_k,z) = \int_{-1}^1dy\ P_\beta (y) \frac{2(1+x_k)} 
{\sqrt{[y+z(1+2x_k)]^2+ 4x_k (x_k+1)(1-z^2)}} .
\label{g-fun}
\end{equation}
For $ \beta = 0 $ the function $g$ actually does not depend on $z$ and is
\begin{equation}
g_0 (x_k) = (1+x_k) \ln \left( 1+\frac{1}{x_k} \right) ,
\end{equation}
so that we recover exactly eq.~(\ref{jt}).

Substituting the result of the loop integration in eq.~(\ref{vel-eps}) 
we finally obtain
\begin{equation}
\epsilon^t_\beta= 2 \sum_k \ {\cal I}_{k1} {\sqrt{x_k}\over 16\pi}
\left[ f_\beta^{(0)} - \frac{\int_{-1}^1 dz \ g_\beta (x_k,z)
P_\beta(z)}{ f_\beta^{(0)}} \right] . 
\label{eps-tb}
\end{equation}
Again we can evaluate this function analytically only for small $\beta$, 
by using the expansion
\begin{equation}
g_\beta (x_k,z) \simeq g_0 (x_k) + \beta z
\frac{M_1/2T}{\sinh(M_1/2T)}  h(x_k) +
\beta^2 \left(\frac{M_1/2T}{\sinh(M_1/2T)}\right)^2 \left[ j(x_k) -  
z^2 l(x_k) \right] ,
\end{equation}
where 
\begin{eqnarray}
h(x) &=& 2(1+x) - (1+2x) g_0(x), \\
j(x) &=& (1+x)(1+2x) -2x(1+x) g_0(x), \\
l(x) &=& 3(1+x)(1+2x) - (1+6x+6x^2) g_0(x).
\end{eqnarray}
So, the average CP violation is given by
\begin{equation}
\langle \epsilon^t \rangle \simeq 2 \sum_k \ 
{\cal I}_{k1}{\rm Im}\!  \left\{J_t(x_k)\right\} \left[ 1 + 
\frac{1}{3} \langle\beta^2  \rangle
\left(\frac{M_1/2T}{\sinh{(M_1/2T)}} \right)^2 
\left( \frac{1}{g_0(x_k) - 1} - 2x_k \right) \right].
\label{epstapp}
\end{equation}

We then see that the effect due to the particle motion is again to increase 
the vertex CP asymmetry, as in the case of the wave part.  However,
 the effect depends 
also on $x_k$, i.e. on the hierarchy between 
the particle masses. In the limit of large $x_k$ we can use the expansion
$(g_0(x_k) - 1)^{-1}\simeq 2 x_k +2/3$, to obtain 
\begin{equation}
\langle \epsilon^t \rangle \simeq \epsilon_0^t \left[ 1 + 
\frac{2}{9} \langle\beta^2  \rangle
\left(\frac{M_1/2T}{\sinh{(M_1/2T)}} \right)^2 \right],
\end{equation}
so that the overall factor coincides with the one in
eq.~(\ref{mean-eps-w}) for $\langle \epsilon^s \rangle $.

To estimate the goodness of the approximate result in
eq.~(\ref{epstapp}) we evaluated the term in square brackets there,
taking $x_k=5$ for definitenes, and we obtained 1.163, 1.055 and 1.000
for $M_1/T=1$, 3 and 10 respectively. A numerical evaluation of the
exact correction due to the velocity, using eq.~(\ref{eps-tb}), leads 
for the same factor the values 1.222, 1.052, 1.000 respectively, so that 
again the accuracy is reasonable, and we see that 
for $T<M_1$ the velocity
dependent correction to $\epsilon$ is not large.

\section{Conclusion}

The baryon asymmetry of the Universe is probably the most important
manifestation  of the existence of $CP$ violation. In the classic
scenarios for baryogenesis through the out of equilibrium decay of
heavy particles in the early stages of rapid expansion of the
Universe, the asymmetries in the $B$ (or $L$) violating decay rates to
conjugate final states arise at the one loop level, involving the
virtual exchange of light particles, such as quarks, leptons or Higgs
bosons. We have studied in this paper the effects of the thermal
background of standard particles present during the decay epoch in the
evaluation of the $CP$ violating asymmetries. We first reconsidered
the triplet scalar decay in $SU(5)$, finding that the asymmetry is
reduced (contrary to an earlier result), as could be expected on the
basis of the Pauli exclusion principle applied to the virtual
fermionic lines in the loop.  We also included the $CP$ violation
produced by the mixing among different heavy states. Confronting with
the $T=0$ results, the modification produced by the thermal effects
could be as large as 50\% for $T\simeq M_1$, but diminishes with
decreasing temperature, becoming negligible for $T<M_1/10$. 

The realization that standard model anomalous $B$ and $L$ violating
processes are in equilibrium at temperatures above the electroweak
phase transition one, has made the $SU(5)$ scenario just of academic
interest, since no net $B-L$ asymmetry (the only one unaffected by
anomalous processes) is generated within it. However, the same
anomalous processes have allowed some new very attractive
possibilities, including the baryogenesis at the electroweak scale
itself (although its practical implementation faces several 
difficulties). The most simple and promising scenario seems to be the
leptogenesis, in which heavy right handed neutrinos generate a lepton
asymmetry in their decay, which is then reprocessed into a baryon
asymmetry by the standard model anomalous processes. We showed that in
these scenarios  the leading thermal corrections canceled among
themselves, due to the opposing effects produced by the bosons and
fermions involved in the loop. Also in the supersymmetric version of
leptogenesis the thermal corrections to the heavy scalar neutrino
decay  were shown to vanish, once the thermal modification of the
branching ratios to the different final states are included. 
In view of this, we studied the correction due to the fact that 
in general the decaying particle is not at rest in the thermal bath.
We showed that the $CP$ asymmetry depends on the particle motion,
since the background density distribution, and hence the thermal
corrections, are now modified in the rest frame of the decaying
particle. This effect can however change the decay asymmetries only by
at most $\sim 20\%$ with respect to the usual $T=0$ results, 
and hence these last can be safely employed.

\begin{center}
\bf{Acknowledgements}
\end{center}

This work was supported in part by CICYT under grant AEN-96/1718 and
DGICYT under grant PB95-1077 (Spain), and 
by EEC under the TMR contract ERBFMRX-CT96-0090.

%\vfill\eject

\pagebreak

{\bf\large Figure captions:}

\bigskip

\noindent Figure 1: Diagrams which interfere with the tree one to produce
the $CP$ violation in the heavy particle decay. Fig.~1a gives the so
called vertex contribution while Fig.~1b gives the  wave function one.

\vfill

\begin{figure}
\centerline{\epsfbox{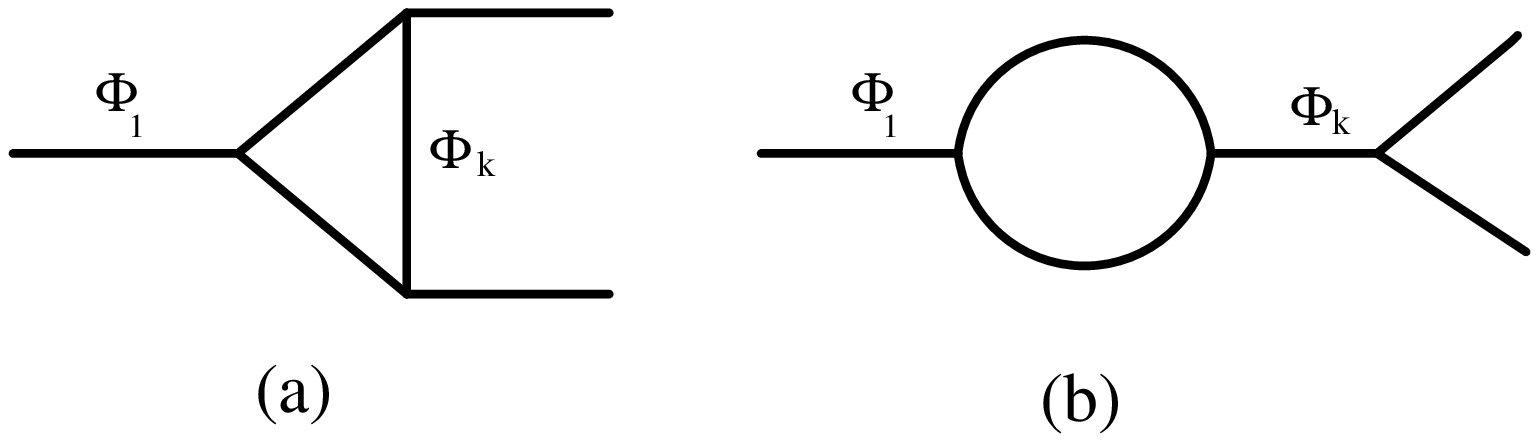}}
\end{figure}


\begin{thebibliography}{100}

\bibitem{ta84} K. Takahashi, Phys. Rev. {\bf D29} (1984) 632.

\bibitem{ig79}A. Yu. Ignatiev, V. A. Kuzmin and M. E. Shaposhnikov,
JETP Lett. {\bf 30} (1979) 688.


\bibitem{bo91}F. J. Botella and J. Roldan, Phys. Rev. {\bf D44} (1991)
966. 

\bibitem{li93} J. Liu and G. Segr\`e, Phys. Rev. {\bf D48} (1993)
4609.


\bibitem{co96} L. Covi, E. Roulet and F. Vissani, Phys. Lett {\bf B384} 
(1996) 169.


\bibitem{li94} J. Liu and G. Segr\`e, Phys. Rev. {\bf D49} (1994)
1342.


\bibitem{co97} L. Covi and E. Roulet, Phys. Lett. B in press,
hep-ph/9611425. 

\bibitem{yi80} A. Yildiz and P. H. Cox, Phys. Rev. {\bf D21} (1980)
906.


\bibitem{na79} D.V. Nanopoulos and S. Weinberg, Phys. Rev. {\bf D20}
(1979) 2484.

\bibitem{rtf} M. Le Bellac, {\it Thermal Field Theory}, Cambridge
University Press, 1996; N. P. Landsman and Ch. G. van Weert, Phys.
Rep. {\bf 145} (1987) 141.

\bibitem{itf} R. Kobes, Phys. Rev. Lett. {\bf 67} (1991) 1384;
M. E. Carrington and U. Heinz, hep-th/9606055.

\bibitem{cutt} R. Kobes and G. W. Semenoff, Nucl. Phys. {\bf B272}
(1986) 329; F. Gelis, hep-ph/9701410.

\bibitem{ma83}N. S. Manton, Phys. Rev. {\bf D28} (1983) 2212;
V. A. Kuzmin, V. A. Rubakov and M. E. Shaposhnikov, Phys. Lett. {\bf B155}
(1985) 36.

\bibitem{fu86} M. Fukugita and T. Yanagida, Phys. Lett. {\bf B174}
(1986) 45.

\bibitem{lgen} M. A. Luty, Phys. Rev. {\bf D45} (1992) 455;
M. Pl\"umacher, hep--ph/9604229; 
 M. Flanz, E. A. Paschos and U. Sarkar, Phys. Lett. {\bf
B345} (1995) 248; Errata: ibidem {\bf B382} (1996) 447,
ibidem {\bf B384} (1996) 487.

\bibitem{ca93} B. A. Campbell, S. Davidson and K. A. Olive,
Nucl. Phys. {\bf B399} (1993) 111.

\bibitem{mu93}H. Murayama, H. Suzuki, T. Yanagida and J. Yokoyama, 
Phys. Rev. Lett. {\bf 70} (1993)
1912.

\bibitem{we82}H. A. Weldon, Phys. Rev. {\bf D26} (1982) 2789.
 

\end{thebibliography}
\end{document}